# The study on quantum material WTe$_2$

Xing-Chen Pan[1], Xuefeng Wang[2], Fengqi Song[1,*] & Baigeng Wang[1]

[1]*National Laboratory of Solid State Microstructures, Collaborative Innovation Center of Advanced Microstructures, and College of Physics, Nanjing University, Nanjing 210093, China*

[2]*National Laboratory of Solid State Microstructures, Collaborative Innovation Center of Advanced Microstructures, and School of Electronic Science and Engineering, Nanjing University, Nanjing 210093, China*

*songfengqi@nju.edu.cn

# The study on quantum material WTe$_2$

WTe2 and its sister alloys have attracted tremendous attentions recent years due to the large non-saturating magnetoresistance and topological non-trivial properties. Herein, we briefly review the electrical property studies on this new quantum material.

**Background**

Recently, tungsten telluride (WTe$_2$), has been recognized as a new rising member of the quantum materials as a platform accommodating many unique physics, like low-dimensionality, strong spin-orbital coupling (SOC) and structural versatility. The atomic structure of WTe$_2$ is totally different from most other MX$_2$ transition metal dichalcogenides (TMDC) 2D materials (1T/2H/3R). The 2H phase (space group P6$_3$/mmc D$^4_{6h}$) MX$_2$ has been widely investigated as a 2D semiconductor with a direct band gap in monolayer [1]. The crystal structure of WTe$_2$ is orthorhombic with space group Pmn2$_1$ (c$^7_{2v}$) [2], called Td phase. Quasi 1D W chains are formed along the a axis, sandwiched between Te layers. The single crystal of WTe$_2$ is structural inversion breaking and non-magnetic. The electronic band structure of WTe$_2$ is also unique in MX$_2$ TMDC 2D materials (most semiconductors). WTe$_2$ is a semimetal with a tiny complicated Fermi surface.

The electrical properties of WTe2 have been studied for about 50 years. In 1966, the Hall coefficient and thermoelectric power of WTe$_2$ have been measured [3]. People also experimental studied WTe$_2$ band structure by Angle-resolved photoemission spectroscopy (ARPES) in 2000 [4]. Both experiments demonstrated the multiband semimetal nature. Among all the TMDC crystals, possessing 2 heaviest elements and semimetallic transport, it was intensively studied till 2014 with the result of non-saturating MR, topological transport and superconductivity. It was even assumed as the possible candidate of the topological superconductors. Herein we briefly review the recent study on crystal $WTe_2$.

**Magnetotransport**

*Non-saturating Titanic ordinary magnetoresistance in WTe2*

The non-saturating Titanic magnetoresistance in WTe2 was demonstrated by M. N. Ali et al. [5]. They reported extremely large MR (XMR) results in samples grown by Br2 transport method: 452,700% at 4.5K in a field of 14.7T, and 13,000,000% at 0.53K in a field of 60T. Large MR was an uncommon property mostly found in magnetic materials. Giant magnetoresistance (GMR) was discovered in thin film structures composed of alternating ferromagnetic and non-magnetic conductive layers [6, 7]. Colossal negative magnetoresistance (CMR) occurs mostly in manganese-based perovskite oxides [8]. In contrast to other large OMR materials like high mobility bismuth or graphite single crystals, there was no saturation of the MR value even at very high fields. As an OMR effect, the MR followed near $H^2$ law from zero to a very high field. In fact, $WTe_2$ is the first material known that displays such a large, non-saturating and parabolic MR.

The XMR ratios vary dramatically in samples with different quality. The XMR can be strongly suppressed in doped or polycrystalline WTe$_2$ [9]. The correlation of sample's residual resistivity ratio (RRR), MR ratio and average mobility can be found by testing series of samples with different qualities [10]. The method used to grow the sample affects the sample quality greatly. The samples grown by Te-flux method have a better performance than the samples grown by chemical vapor transport (CVT) method. During the flux growth, the more slowly the crystals are cooled, the better the quality is. Among the samples grown by CVT methods, different transport agents used in growing processes bring different results: Br$_2$ performs better than TeBr$_4$, generally. WTe2 exhibits 3-dimensional electrical transport although it is a layered compound with metal layers (W) sandwiched between adjacent insulating chalcogenide layers (Se). L. R. Thoutam et al. [11] found a scaling behavior of the resistance. The mass anisotropy γ is quite limited and as low as 2 at higher temperatures. Under low temperature the mass anisotropy γ varied with temperature and followed the magnetoresistance behavior of the Fermi liquid state. Y. Zhao et al. [12] studied the angular dependence of the magnetoresistance in a high-quality flux-growth WTe$_2$ single crystal. Unexpectedly, when the applied field and excitation current were both parallel to the tungsten chains of WTe$_2$, an exotic large longitudinal linear magnetoresistance as high as 1,200% at 15 T and 2 K was identified. The anisotropic magnetoresistance is attributed to the 1-dimensional transport while the field is aligned precisely along the one-dimensional W chain.

The temperature dependence of MR [5] and Hall coefficient [13] demonstrate an interesting "turn on" effect. The "turn on" temperature of MR was found to be linear with the field at high field region. By fitting the Hall resistivity to the two band model, Y. Luo et al. [14] studied the temperature dependence of carrier density of $WTe_2$. A sudden increase of the hole density below ~160 K and a more pronounced reduction in electron density below 50K were observed. Only at low temperature, the electron and hole densities were found comparable. According to the semi-classical magnetotransport theory [5, 15], the breaking of the electron-hole compensation should be responsible to the MR "turn on" effect. Y. L. Wang [16] et al. found that the MR follows the Kohler's rule. A scaling law of field dependence of "turn on" temperature was also obtained.

*Fermi surface studies*

The mechanism of this unexpected XMR has attracted lots of attention. According to the semi-classical magnetotransport theory (two band model), the conductivity tensor σ can be expressed in the complex representation:

$$\hat{\sigma} = e \left[ \frac{n\mu}{(1+i\mu B)} + \frac{p\mu'}{(1-i\mu' B)} \right]$$

where e>0 is the charge, and μ and μ' are the mobilities of electrons and holes, respectively. The complex resistivity ρ is the reciprocal of σ:

$$\hat{\rho} = \frac{1 + \mu\mu' B^2 + i(\mu - \mu')B}{e(n\mu + p\mu' + i(p-n)\mu\mu' B)}$$

From this equation we know, if p=n, we can observe the non-saturating parabolic MR. The XMR is reasonable if perfect electron-hole compensation exists in WTe$_2$. The fermiology study is thus necessary in solving this MR mystery.

*ARPES*

ARPES technique has been widely used to study the band structures of strong correlated [17] and topological [18] materials.

I. Pletikosić et al. [19] first studied the fermiology of WTe$_2$ at different temperatures. 4 Fermi pockets were observed along the Γ-X direction, 2 from hole and 2 from electron. At low temperatures, Hole and electron pockets of approximately the same size were found. A change in the Fermi surface with temperature has also been discovered: the temperature increase caused the bands forming the pockets to become electron doped. A temperature-induced Lifshitz transition at T ≃ 160 K was observed by Y. Wu et al. [20] with ultrahigh resolution ARPES.

Further experiments present the fermiology of WTe$_2$ in higher resolution, where at least 9 pockets were discovered [13, 21-23]. This includes 1 hole pocket at the Brillouin zone centre Γ, and 2 hole pockets and 2 electron pockets on each side of Γ along the Γ-X direction.

Strong spin-orbital coupling (SOC) effect was demonstrated by using ARPES technique. J. Jiang et al. [21] first combined the circular dichroism (CD) ARPES and density-functional-theory calculations in the study of WTe$_2$. The strong CD effect indicated an exotic spin structure of this material. The spin-texture was determined by spin-resolved ARPES [22, 23]. The spin polarization changed sign upon crossing the

Brillouin zone centre. The time reversal symmetry must be preserved in WTe$_2$, and we can exclude the magnetic origin. The lifting of the spin degeneracy may origin from the non-centrosymmetric crystal structure and strong SOC of WTe$_2$. Using scanning tunnelling microscopy (STM) technique, P. K. Das et al. also found that the electron-hole compensation only exists in at least 3 Te-W-Te layers [23].

*Quantum oscillations*

Quantum oscillation experiments [20, 24-26] was carried out, revealing a series of Fermi pockets in WTe$_2$. P. L. Cai [24] et al. studied the Shubnikov–de Haas oscillation (SdHO) effect under high pressure. 4 pockets were identified, and 2 were found to persist to high pressure. The sizes of these two pockets were comparable, but showed increasing difference with pressure. The drastic decrease of MR under high pressure was attributed to the change of Fermi surfaces.

Z. Zhu et al. [25] studied the thermal power under high field by measuring Nernst and Seebeck coefficients. The Nernst signal was linear in magnetic field under low temperature. By analysing the quantum oscillations observed at different angles, Fermi surfaces were determined. After mapping the Fermi surfaces, the total electron and hole density can be obtained: n=6.44×10$^{19}$cm$^{-3}$, p=6.9×10$^{19}$cm$^{-3}$.

*Electron-hole balance: a mobility spectrum study*

Solving the density of total electron and hole carriers by analysing Hall effect directly is the most convincing demonstration for semi-classical magnetoresistance theory [15]. Traditionally, for a sample involving more than one channel/type of carriers, the conductivity tensor components can be fit as a sum over the m pockets present within the multi-carrier fitting system [14, 26]. In WTe2 with unknown number of Fermi pocket, the mobility spectrum [13] can be used to address this problem.

By calculating the normalized conductivity matrix of the electron- and hole-like carriers, the mobility spectra at various temperatures can be obtained [13]. Although $WTe_2$ has a complicated Fermi surface with lots of pockets, a single effective carrier component was clearly identified for the electron and hole sides at all temperatures considered. The carrier density at various temperatures can be obtained from the integrals of the mobility spectra for the data. This result supported theoretical predictions that an exact compensation (n=p) at low temperatures leads to a cancellation of the Hall-induced electrical field and contributes to the non-saturating parabolic MR [5].

*Magnetoresistance in devices and gate tuning*

A simple backgate can break the carrier compensation and thus check the above carrier-balance physics. The $WTe_2$ nanodevices were fabricated by the mechanical exfoliation method from the bulk single crystal. L. Wang et al. [27] studied the layer dependence of $WTe_2$ transport properties. Breaking of the compensation and metal-insulator transition were found in a few monolayers WTe2 devices. The gate voltage dependence of magnetoresistance was also studied. They also [28] found a long- range field effect in $WTe_2$ devices, leading to large gate induced changes of transport through crystals much thicker than the electrostatic screening length, without changing bulk state much. Y. Wang et al. [29] broke the electron-hole compensation in $WTe_2$ devices. However, the MR keeps still nonsaturating in their experiments. V. Fatemi et al. [30] reported an experiment where both the magnetoresistance and the carrier ratio can be tuned in their metallic nanodevice. The magnetoresistance power law was found subquadratic and independent to the carrier density.

**Phase transition under pressure**

*Pressure induced superconductivity*

The density of states at the Fermi level ($N(E_F)$) is rather low [4] and no superconductivity has ever been expected even down to 0.3 K [24] in $WTe_2$. High pressure is applied in the material, which increases the $N(E_F)$ and drives the superconductivity. Furthermore, in $WTe_2$, the Te-5p and W-5d orbitals are spatially extended, thus very sensitive to the variations caused by the external pressure and strain. The pressure induced superconductivity has been observed by X.-C. Pan et al. [31] and D. Kang et al. [32]. Both experiments observed a dome-shaped superconductivity $T_C$-P phase diagram and a strong MR suppression under high pressure. The detail of the superconductivity phase diagram seemed to be sample dependent. X.-C. Pan et al. [31] observed that the superconductivity sharply appeared at 2.5 GPa, quickly reaching a maximum critical temperature of 7 K at around 16.8 GPa, and followed by a monotonic decrease in $T_C$ with increasing pressure exhibiting the typical dome-shaped superconducting phase. The superconductivity observed by D. Kang et al. [32] emerged at 10.5 GPa with a sign change of Hall coefficient. The origin of the superconductivity also has been discussed. From the theoretical calculations, X.-C. Pan et al. [31] interpreted the low pressure region of the superconducting dome to an enrichment of the density of states at the Fermi level ($N(E_F)$) and attributed the high-pressure-region decrease of the $T_C$ to a possible structural instability. D. Kang et al. [32] attributed the superconductivity to a possible quantum phase transition because of the observed sign change of Hall coefficient. In contrast to the result obtained under ambient pressure, X. -C. Pan et al. [13] observed linear MR effect under high pressure with field along the c axis.

*Structural phase transition under high pressure*

To determine the structural phase under phase is very important to study the electrical properties of WTe$_2$. First structural data under high pressure was given by D. Kang et al. [32]: they conducted a pressure-dependent synchrotron XRD in WTe$_2$ at a wavelength of 0.6199 Å and to a pressure of 20.1 GPa and claimed no structural transition. However, by using the synchrotron XRD with a smaller wavelength of 0.4246 Å and therefore in an enlarged 2θ range as well as extending the highest pressure up to 33.8 GPa, Y. Zhou et al. [33] observed a pressure-induced Td to 1T' (space group of P2$_1$/m) structural transition in WTe$_2$. The structural phase transition started at 6 GPa and completed above 15.5 GPa. The structural transition also casted a pressure range with the broadened superconducting transition, where the zero resistance disappeared. This Td to 1T' phase transition was also investigated by P. Lu et al. [34]. With ab initio calculations and high-pressure synchrotron x-ray diffraction and Raman spectroscopy, they found a Td to 1T' phase transition at around 4-5 GPa. Moreover, they attributed the emerging of the superconductivity to this phase transition. Interestingly, the dome-shaped superconductivity phase diagram can also be observed in the sister compound of WTe$_2$, MoTe$_2$ [35], showing a 1T' to Td structural phase transition when cooling down the temperature.

**Topological non-trivial state in WTe$_2$**

*Type-II Weyl semimetal states*

Weyl fermions predicted by H. Weyl [36] may be realized as a quasiparticle in solid materials [37, 38] . Thanks to the progress of ARPES technique, people have observed the quasiparticle topological excitation of Weyl fermions in monophosphide [39, 40] . Unique characters include several couples of Weyl cones with linear dispersion in bulk electronic structure and A surface state, i.e. Fermi arc [38] connecting a couple of Weyl points (WPs) with opposite chirality. The chiral anomaly can be detected by anisotropic negative magnetoresistance measurements [41].

The normal Weyl semimetals have a point-like Fermi surface, i.e. WP. Recently, the type-II WPs is predicted, which exist at the boundaries between electron and hole pockets [42]. There're several couples of highly tilted Weyl cones in bulk electronic structure. The Fermi arc remains in the surface of the type-II Weyl semimetal. However, the chiral anomaly effect [42] and quantum oscillation [43-45] may be more complicated. The first predicted type-II Weyl semimetal candidate is WTe$_2$[42]. Her sister compound, MoTe$_2$ was later proposed to be a type-II Weyl semimetal [46-48], which is demonstrated by ARPES soon [49-51]. TaIrTe$_4$ [52] and LaAlGe [53] were also demonstrated to hold type-II Weyl fermions. Recently, the band structure with highly tilted cone has also been found in type-II Dirac semimetals [54-58].

*Fermi arc in WTe$_2$ and Mo$_x$W$_{1-x}$Te$_2$*

As the first predicted type-II topological Weyl semimetal, intense efforts have been made in ARPES measurements of WTe$_2$[59-62]. The WP is 50meV over the Fermi level, therefore, most ARPES efforts are focused on the Fermi arc. The problem is whether the observed surface state is topological non-trivial or not. The first ARPES employ the laser pump technique to probe the fermi surface over the fermi level, where a surface arc is obtained, while it was later found topologically trivial. C. Wang et al. [59] claimed that the observed Fermi arc was consistent with the type-II Weyl nature. Y. Wu et al. [60] found that different domains in the same sample showing different topological classifications. F. Y. Bruno et al. [61] and J. Sanchez-Barriga et al. [62] also observed topologically trivial Fermi arc since the observed arc did not connect the electron and hole pockets. The bulk and surface correspondence of WTe$_2$ was found more similar to a topological insulator rather than a type-II Weyl semimetal.

The length of the predicted topological Fermi arc is too small to clearly resolve by the present APRES technique in WTe$_2$. T.-R. Chang et al. [63] find that Mo doping may elongate the Fermi arc. In other words, Mo$_x$W$_{1-x}$Te$_2$ holds an arc-tunable Weyl fermion metallic state. This is realized in the experiments of I. Belopolski et al. [64], where a series ratio of Mo is introduced in WTe$_2$. The ARPES mapping gives a Mo-ratio dependent fermi Arc, The Chern number of the system can be counted according to the theoretically analysis. By comparing results with ab initio calculations, the topological Fermi arc was demonstrated in Mo$_{0.45}$W$_{0.55}$Te$_2$.

*Approaching the Weyl points*

The predicted Weyl cones of $WTe_2$ is high tilted, hence only when the Fermi level near WPs, the low energy topological excitation can be detected via transport measurements. Y. Wang et al. [65] tuned the Fermi level through the WPs via the electric field effect in a WTe2 thin film device. A negative longitude magnetoresistance was observed, and was attributed to the type-II topological Weyl nature. Y.-Y. Lv et al. [66] also observed the significant anisotropic Adler-Bell-Jackiw (ABJ) anomaly and possible temperature induced topological phase transition in the Fermi-level delicately adjusted $WTe_{1.98}$ crystals. Similar negative magnetoresistance effect has also been observed in $WTe_2$ nanosheets synthesised by chemical vapor deposition technique [67]. However, this chiral anomaly effect hasn't been found its sister compound $MoTe_2$ till now.

Finding a Fermi level tunable system by band structure engineering [68, 69] is another effective way in topological material research. I. Belopolski et al. [70] demonstrated in $Mo_xW_{1-x}Te_2$, that a group of WPs were at the Fermi level, making $Mo_xW_{1-x}Te_2$ a promising platform for transport and optics experiments on Weyl semimetals. H. Zheng et al. [71] visualized the interference pattern on $Mo_{0.66}W_{0.34}Te_2$ by atomic resolution scanning tunnelling microscopy and spectroscopy (STM-STS). The quasiparticle pattern (QPI) observed was found dominated by surface states.

Recently, I. Belopolski et al. [72] observed a topological Weyl phase transition in $Mo_xW_{1-x}Te_2$ system. Bulk $WTe_2$ state was found to be topological trivial, and $Mo_xW_{1-x}Te_2$ (x>0.07) was non-trivial, according to the topological phase diagram obtained. For the first time, the creation of magnetic monopoles in momentum space was demonstrated in this system.

*2D topological insulator state in WTe2 monolayer*

The experimental realization of 2D topological insulator and quantum spin Hall (QSH) effect lifted the curtain on recent 10 years' topological materials studies. QSH systems or topological insulators have an energy band gap separating the valance and conduction bands in the bulk. However, they have gapless edge states protected by time-reversal symmetry on the boundary [73]. Contrast to the quantum Hall (QH) effect, the spinful 1D chain on the boundary in QSH does not require applied magnetic field. X. Qian et al. [74] predicted a class of large-gap QSH insulators in two-dimensional transition metal dichalcogenide monolayers with 1T' structure, namely, 1T'-$MX_2$ with M= (Mo or W) and X= (S, Se or Te) by using first-principles calculations. The result indicated $WTe_2$ monolayer to be a 2D topological insulator candidate.

Several groups have studied this topological property of $WTe_2$ monolayer. Z. Fei et al. [75] performed low-temperature electrical transport experiment on $WTe_2$ devices. Insulating state was found in monolayer and bilayer $WTe_2$ devices. In the monolayer device, edge conductance was believed to appear. However, the quantized conductance couldn't be seen. Electronic structure of monolayer $WTe_2$ was discovered by S. Tang et al. [76] and Z.-Y. Jia et al. [77]. Topological band inversion and band gap opening were observed by ARPES in monolayer WTe2 films grown by molecular beam epitaxy (MBE). An edge state was confirmed in the insulating monolayers by STM.

**Conclusion & Outlook**

Owing to the tiny compensate Fermi surface and high electrical mobility, $WTe_2$ is the first material known that displays an anisotropic non-saturating titanic MR. Superconductivity phase transition and Strong SOC effects have also been observed. Thanks to these excellent electrical properties, $WTe_2$ is a quantum material of massive potentialities in electronic and spintronic applications. It may be useful in devices such as highly sensitive low-temperature magnetic-field sensors or high-field temperature sensors in cryogenics [5]. Controlling of spin–orbit torques has been realized in WTe2/Py bilayers by D. MacNeill et al. [78], where WTe2 acted as a spin-source. Topological metal properties in WTe2 and its sister compounds have been studied by several groups. As the first type-II Weyl semimetal candidate, the topological classification of WTe2 bulk state is still controversial, due to different results from several experiments. In monolayer WTe2, signature of QSH effect has been observed. As the first layered 2D topological insulator candidate, WTe2 has a fascinating prospect in condensed matter physics studies.

This work was supported by the National Key Projects for Basic Research of China (Grants No. 2013CB922103 and No. 2017YFA0303200), the National Natural Science Foundation of China (Grants No. 91421109, No. 91622115, No. 11522432, No. 61176088 and No. 11274003), the Natural Science Foundation of Jiangsu Province (Grants No. BK20130054 and No. BK20160659) and the Fundamental Research Funds for the Central Universities (020414380056 and 020414380057). Research funding

from the education ministry is also acknowledge. Xing-Chen Pan would like to acknowledge the helpful discussions with Juan Jiang, Huimei Liu, Yiming Pan and Ilya Belopolski.

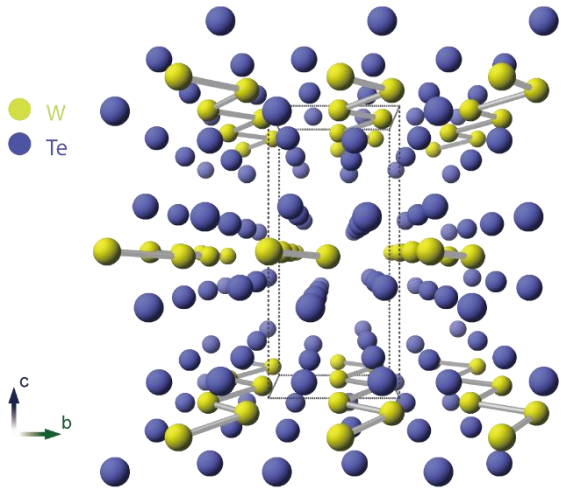

Figure.1 The atomic structure and crystal data for WTe$_2$. From Ref. [31].

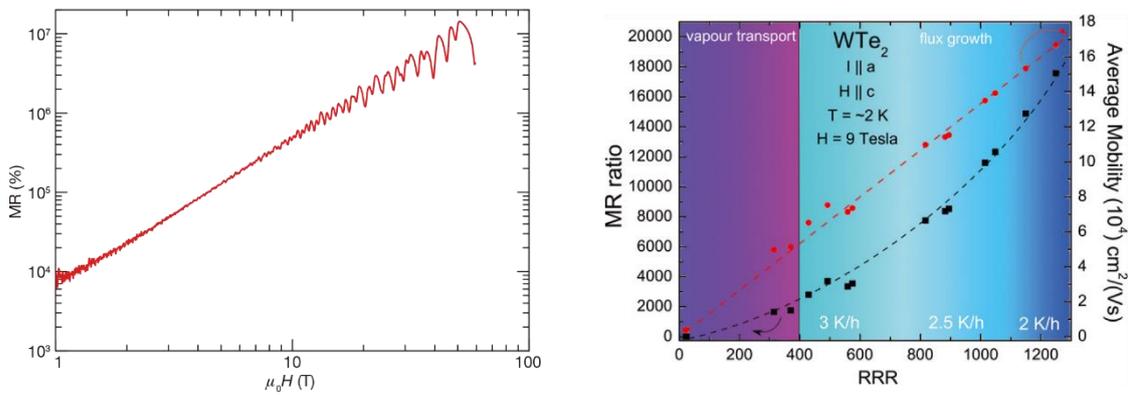

Figure.2 XMR in WTe$_2$ and the correlation of the MR ratio with the RRR value.

From Ref. [5, 10].

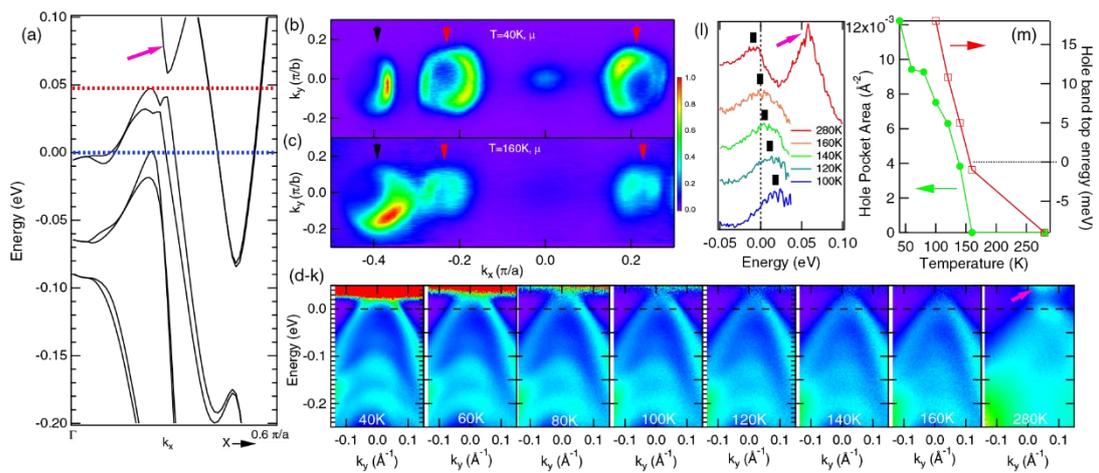

Figure.3 Temperature-induced Lifshitz transition in WTe$_2$. From Ref. [20].

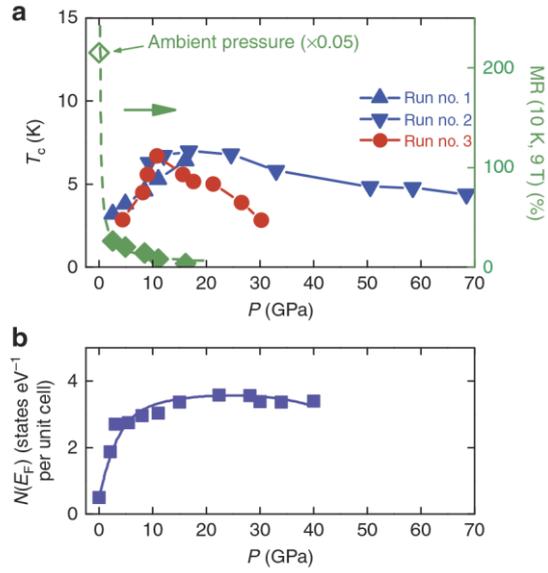

Figure.4 Superconducting $T_C$-P phase diagram of $WTe_2$. From Ref. [31].

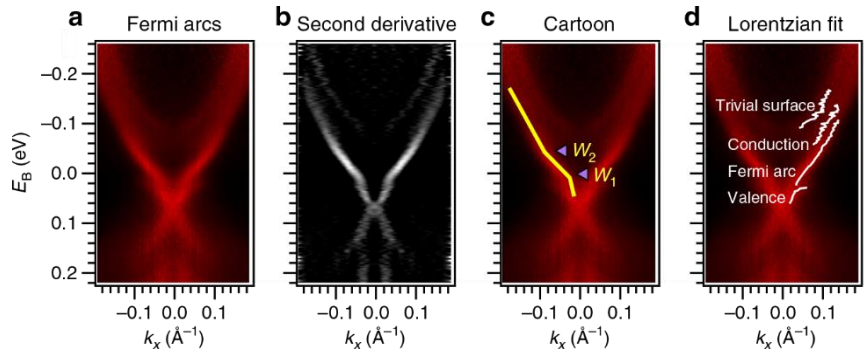

Figure.5 Fermi arcs in $Mo_xW_{1-x}Te_2$. From Ref.[70].

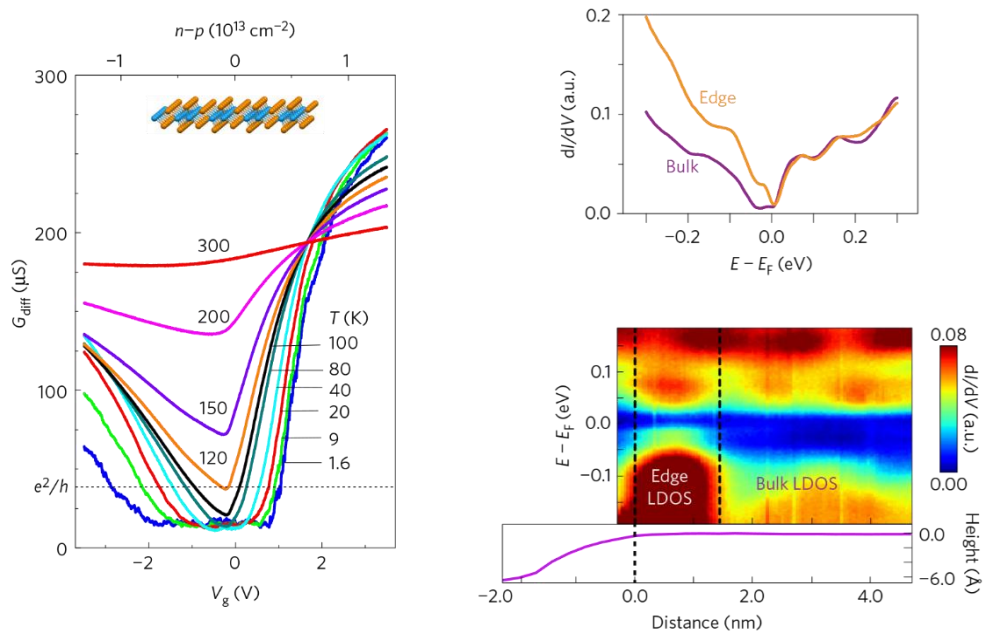

Figure.6 Topological edge states in monolayer $WTe_2$. From Ref. [75, 76].